\documentclass[12pt, titlepage]{article}
\usepackage{authblk}
\usepackage[margin=1in]{geometry}
\RequirePackage{amsthm,amsmath,amsfonts}
\RequirePackage{natbib}
\RequirePackage[colorlinks,citecolor=blue,urlcolor=blue]{hyperref}
\usepackage{booktabs,longtable}
\usepackage{setspace}
\usepackage{graphicx}
\usepackage{xcolor}
\usepackage{epstopdf}
\usepackage{tikz}
\usetikzlibrary{arrows.meta}
\usepackage{subcaption}
\graphicspath{{./}{../result/figures}}
\usepackage{bm}
\usepackage[ruled, lined]{algorithm2e}

\allowdisplaybreaks

\usepackage[pagewise]{lineno}
\newcommand*\patchAmsMathEnvironmentForLineno[1]{%
        \expandafter\let\csname old#1\expandafter\endcsname\csname 
        #1\endcsname
        \expandafter\let\csname oldend#1\expandafter\endcsname\csname 
        end#1\endcsname
        \renewenvironment{#1}%
        {\linenomath\csname old#1\endcsname}%
        {\csname oldend#1\endcsname\endlinenomath}}%
\newcommand*\patchBothAmsMathEnvironmentsForLineno[1]{%
        \patchAmsMathEnvironmentForLineno{#1}%
        \patchAmsMathEnvironmentForLineno{#1*}}%
\AtBeginDocument{%
        \patchBothAmsMathEnvironmentsForLineno{equation}%
        \patchBothAmsMathEnvironmentsForLineno{align}%
        \patchBothAmsMathEnvironmentsForLineno{flalign}%
        \patchBothAmsMathEnvironmentsForLineno{alignat}%
        \patchBothAmsMathEnvironmentsForLineno{gather}%
        \patchBothAmsMathEnvironmentsForLineno{multline}%
}

\let\proglang=\textsf 

\newcommand{\bmW}{\bm{W}}
\newcommand{\bmX}{\bm{X}}
\newcommand{\bmS}{\bm{S}}
\newcommand{\bmC}{\bm{C}}

\DeclareMathOperator*{\argmax}{arg\,max}

\title{Covariate Connectivity Combined Clustering for Weighted Networks}

\author[1]{Zeyu Hu}
\author[1]{Wenrui Li}
\author[1]{Jun Yan}
\author[2,*]{Panpan Zhang}

\affil[1]{Department of Statistics, University of Connecticut, 
Storrs, CT 06269}
\affil[1]{Department of Biostatistics, Vanderbilt University Medical 
Center, Nashville, TN 37203}
\affil[*]{Correspondence: \href{mailto:panpan.zhang@vumc.org}{Panpan 
Zhang}}

\begin{document}
\doublespacing
\maketitle

\begin{abstract}
Community detection is a central task in network analysis, with
applications in social, biological, and technological
systems. Traditional algorithms rely primarily on network topology,
which can fail when community signals are partly encoded in
node-specific attributes. Existing covariate-assisted methods often
assume the number of clusters is known, involve computationally
intensive inference, or are not designed for weighted networks. We
propose $\text{C}^4$: Covariate Connectivity Combined Clustering, an 
adaptive spectral clustering algorithm that integrates network 
connectivity and node-level covariates into a unified similarity 
representation. $\text{C}^4$ balances the two sources
of information through a data-driven tuning parameter, estimates the
number of communities via an eigengap heuristic, and avoids reliance
on costly sampling-based procedures. Simulation studies show that 
$\text{C}^4$ achieves higher accuracy and robustness than competing 
approaches across diverse scenarios. Application to an airport 
reachability network demonstrates the method's scalability, 
interpretability, and practical utility for real-world weighted 
networks.

\bigskip

\noindent{\bf Key words.}
 community detection; covariate-assisted learning; network analysis;
 spectral clustering; similarity fusion; tuning parameter selection
\end{abstract}

\section{Introduction}
\label{sec:intro}

Community detection is a central problem in network analysis that 
seeks to identify groups of nodes that are densely interconnected or 
share similar attributes within a larger
network~\citep{girvan2002community,
  fortunato2010community}. Uncovering such communities reveals the
organizational structures or functional modules that govern complex 
systems. In social networks, community detection has been widely 
used to analyze collaboration and interaction
patterns~\citep{barabasi2002evolution, handcock2007model, ji2022co}, 
with methodological advances further enhancing the modeling of such 
structures~\citep{ouyang2021clique, ouyang2023mixed}. In biological 
systems, modular organization plays a fundamental role in processes 
such as brain function~\citep{sporns2016modular}. In
technological and information networks, community detection supports
the analysis of incomplete or large-scale
structures~\citep{lin2012community}. Applications also extend to
public health, where clustering methods have been employed to group
states by disease dynamics and to model covariate-assisted network
structures~\citep{chen2022clustering, louit2025calfsbm}. The ubiquity
of network data across these diverse domains underscores the 
importance of developing reliable community detection methods to 
uncover hidden structures and gain insights into the organization and 
function of real-world systems.

Classical community detection methods primarily rely on network 
topology. Modularity optimization approaches, including the 
Girvan--Newman algorithm and its
successors~\citep{girvan2002community, newman2004finding, 
newman2006modularity}, remain influential for their ability to reveal 
cohesive modules in networks,
with the Louvain method especially popular for its efficiency and
scalability on large graphs~\citep{blondel2008fast}.
Spectral clustering methods exploit the 
eigenspectrum of graph Laplacians to identify low-dimensional 
embeddings of network structure~\citep{shi2000normalized, 
ng2002spectral, vonluxburg2007tutorial}, while other graph-based 
strategies include edge betweenness~\citep{radicchi2004defining}, 
random-walk methods~\citep{pons2005computing, rosvall2008maps}, and 
label propagation~\citep{raghavan2007near}. Probabilistic 
formulations such as the mixed-membership stochastic block model 
extend this topological perspective~\citep{airoldi2008mixed}. These 
methods effectively detect well-separated clusters but struggle when 
relevant signals are not fully encoded in connectivity alone, 
motivating the use of node attributes to enhance community detection.

Covariate-assisted methods extend community detection by combining 
network topology with node attributes. Covariate-assisted spectral
clustering (CASC) integrates connectivity and attributes but requires
the number of communities to be specified in
advance~\citep{binkiewicz2017covariate}. To improve
performance in sparse networks, regularization strategies have been 
proposed to stabilize estimation~\citep{yan2021covariate}.
Bayesian frameworks offer complementary strategies. 
\citet{tallberg2004bayesian} introduced one of the earliest 
covariate-dependent block models, and~\citet{newman2016structure} 
formalized inference in annotated networks. Recent developments 
extend the stochastic block model to incorporate covariates via 
random partition models~\citep{shen2025bayesian}, latent factor
formulations~\citep{louit2025calfsbm}, and multilayer
settings~\citep{contisciani2020community, xu2023covariate}. Empirical 
applications also demonstrate the utility of attributes, such as in 
collaboration networks of statisticians~\citep{zhang2023community}. 
Despite these advances, no existing method efficiently integrates 
weighted network structure with informative covariates while jointly 
balancing the two information sources and determining the number of 
communities in a fully data-driven manner.

To fill this gap, we propose Covariate Connectivity Combined 
Clustering ($\text{C}^4$), a simple yet effective algorithm that fuses network 
connectivity with node covariates. The method jointly incorporates 
weighted network structure and informative node attributes, 
adaptively selects the tuning parameter that balances these two 
sources of information, and determines the number of communities in a 
data-driven manner. In contrast to existing Bayesian formulations, 
$\text{C}^4$ avoids reliance on computationally intensive Markov chain Monte 
Carlo procedures and remains highly scalable. Through extensive 
simulations, we show that $\text{C}^4$ achieves higher clustering accuracy and 
robustness than competing methods across a wide range of network 
configurations, both when the number of communities is known and 
unknown. Application to an airport reachability network reveals 
geographically coherent and demographically meaningful communities, 
further demonstrating its scalability and interpretability,. 
Together, these results establish $\text{C}^4$ as a practical and 
robust solution for covariate-assisted community detection in 
weighted networks.

The remainder of the paper is organized as follows. 
Section~\ref{sec:method} details the proposed $\text{C}^4$ 
algorithm, including the construction of the covariate-based 
similarity matrix, the fusion of connectivity and covariate 
information, the selection of the tuning parameter, and the 
practical implementation. Section~\ref{sec:sim} reports simulation 
studies conducted under both known and unknown community numbers, 
evaluating the accuracy and robustness of $\text{C}^4$ across a 
variety of network configurations. 
Section~\ref{sec:real_data} presents an application to an airport 
reachability network, illustrating the scalability and 
interpretability of the method in practice. Section~\ref{sec:disc} 
discusses the limitations and future directions.

\section{Method}
\label{sec:method}

We begin with introducing notations for weighted networks and 
covariates (Section~\ref{sec:notation}). Then, we detail how 
connectivity and covariate information are fused within a spectral 
clustering framework (Section~\ref{sec:c4}) and describe the 
procedure for selecting the tuning parameter 
(Section~\ref{sec:tuning}). Implementation details
are provided in Section~\ref{sec:impl}.

\subsection{Notations}
\label{sec:notation}

We first introduce notations for the weighted network and node-level
covariates. Let $G := G(V, E)$ represent a weighted, undirected 
network with node set~$V$ and edge set~$E$.
Let $|V| = n$ be the number of nodes in~$G$.
The structure of $G$ is characterized by a 
weighted adjacency matrix $\bmW := (w_{ij})_{n \times n}$, where 
$w_{ij}$ is the weight between nodes $i, j \in V$. If nodes~$i$
and~$j$ are not connected, then we set $w_{ij} = 0$. Further, we
do not consider self-loops by enforcing $w_{ii} = 0$ for all
$i \in V$. For each node $i$, let $z_i \in \{1, 2, \ldots, K\}$ 
denote its community label, where $K$ is the total number of 
communities. Collectively, $\bm{z} = (z_1, z_2, \ldots, z_n)^{\top}$ 
represents the clustering assignment of all nodes. In practice, $K$ 
is unknown.

Each node~$i$ is endowed with a $p$-dimensional covariate
vector $\bmX_i$ collecting the node-level features. Let
$f := \mathbb{R}^p \times \mathbb{R}^p \mapsto \mathbb{R}$
be a similarity function that quantifies the closeness between
$\bmX_i$ and $\bmX_j$ for each pair of $i, j \in V$, with larger
values indicating greater similarity. Distances followed by
inverse transformations are commonly used: Euclidean distance for
continuous variables, Hamming distance for categorical variables,
and more flexible metrics such as Gower’s distance for mixed
data~\citep{wang2021hybrid}. The choice of inverse transformation
depends on both the selected distance function and the practical 
interpretation of the real data. Let
$\bmS := (s_{ij})_{n \times n}$ denote the similarity matrix, with
$s_{ij} = f(\bmX_i, \bmX_j)$.

\subsection{Covariate Connectivity Combined Clustering}
\label{sec:c4}

Our $\text{C}^4$ integrates network topology and node-level 
covariates within a spectral clustering framework. This methods 
enables uncovering communities that are not only structurally 
cohesive but also homogeneous with respect to relevant nodal 
attributes. Define
$\bmC := (c_{ij})_{n \times n} = (1 - \alpha) \bmW + \alpha \bmS$,
where 
$\alpha \in [0, 1]$ is a tuning parameter balancing the structural and
covariate similarity. Setting $\alpha = 0$ reduces to structure-only
clustering, while $\alpha = 1$ refers to covariate-only
clustering. Intermediate values of $\alpha$ fuse the two information
sources. Because $\bm{W}$ and $\bm{S}$ may differ in scale, we rescale
$\bm{S}$ so that its total sum matches that of $\bm{W}$:
\begin{equation}
  \label{eq:rescale}
  \bm{S} \leftarrow \bm{S} \times
  \frac{\sum_{i,j} w_{ij}}{\sum_{i,j} s_{ij}}.
\end{equation}
This adjustment ensures that neither source dominates due to
magnitude differences.

For any fixed choice of~$\alpha$, spectral clustering is performed on
the normalized Laplacian of $\bmC$:
\begin{equation}
  \label{eq:nlaplacian}
  \bm{L}_{\text{norm}} = \bm{I} - \bm{D}^{-1/2} \bm{C} \bm{D}^{-1/2},
\end{equation}
where $\bm{I}$ is the identity matrix, and $\bm{D}$ is a diagonal 
matrix with $d_{ii} := \sum_{j = 1}^{n} c_{ij}$.  Spectral 
clustering interprets
community detection as an eigenvalue problem, embedding nodes in the
eigenspace of $\bm{L}_{\text{norm}}$~\citep{chung1997spectral,
  ng2002spectral, vonluxburg2007tutorial}. Communities are then
identified using $k$-means applied to the eigenvectors corresponding 
to the smallest eigenvalues. The number of clusters $K$ is chosen 
adaptively via an eigengap heuristic~\citep{ng2002spectral}. 
Specifically, if $\lambda_{(1)}, \lambda_{(2)}, \ldots, 
\lambda_{(n)}$ are the ordered eigenvalues of
$\bm{L}_{\text{norm}}$, we set
\begin{equation}
  \label{eq:opt_K}
  K = \arg\max_{k \in \{2,\ldots,n-1\}}
  \big(\lambda_{(k+1)} - \lambda_{(k)}\big).
\end{equation}
The multiplicity of the zero eigenvalue of $\bm{L}_{\text{norm}}$
equals the number of connected components in the graph. In practice,
when the graph is connected, there is a single zero eigenvalue; the
eigengap heuristic is applied to the ordered sequence of eigenvalues,
including the zero(s).

\subsection{Tuning Parameter Selection}
\label{sec:tuning}

The tuning parameter $\alpha$ is selected through a data-driven grid
search that maximizes the silhouette score, a distance-based 
criterion measuring how well nodes are clustered. Since its
appearance~\citep{rousseeuw1987silhouettes}, the silhouette
score has become a standard internal 
validation criterion because it simultaneously quantifies 
within-cluster cohesion and between-cluster separation, while not 
requiring external labels~\citep{shahapure2020cluster}. Because of 
its robustness, direct interpretability, and ease of use, the 
silhouette score has become a prevalent metric for model selection, 
particularly in clustering research~\citep{dudek2020silhouette, 
januzaj2023determining}.

To operationalize this selection, the silhouette score is computed
from pairwise node distances derived from the fused adjacency-similarity
matrix~$\bm{C}$. The distance between nodes~$i$ and~$j$ is defined as
$d(i,j) = 1/(c_{ij} + \varepsilon)$, where $c_{ij}$ denotes the
integrated similarity in~$\bm{C}$ and $\varepsilon > 0$ provides
numerical stability. This definition ensures that stronger
similarities correspond to shorter distances. For each
candidate~$\alpha$ in the pre-specified grid $\mathbb{A}$, we
construct~$\bm{C}$, estimate the number of clusters $K$ by the
eigengap heuristic (Section~\ref{sec:c4}), and evaluate the silhouette
score. The final choice of both $\alpha$ and $K$ are obtained
jointly by selecting the pair $(\alpha, K)$ that maximizes the
silhouette score.

\subsection{Implementation}
\label{sec:impl}

\begin{algorithm}[tbp]
\SetArgSty{textrm}
\SetAlgoLined
\LinesNumbered
\caption{Pseudocode for $\text{C}^4$.}
\label{alg:c4}
\KwIn{Weighted adjacency matrix $\bm{W}$;
  covariate similarity matrix $\bm{S}$;
  grid for tuning parameter $\mathbb{A}$;
  number of clusters $K_{\text{true}}$ (optional).}
\KwOut{Community labels $\bm{z}$;
  optimal tuning parameter $\alpha_{\text{opt}}$;
  number of clusters $K_{\text{opt}}$ (user-provided or data-driven).}
\BlankLine
Rescale $\bm{S}$ with Equation~\eqref{eq:rescale}\;
\ForEach{$\alpha \in \mathbb{A}$}{
    Construct $\bm{C} \leftarrow (1 - \alpha)\bm{W} + \alpha \bm{S}$\;
    Compute the normalized Laplacian $\bm{L}_{\text{norm}}$ of 
    $\bmC$ using Equation~\eqref{eq:nlaplacian}\;
    Calculate and sort the eigenvalues of $\bm{L}_{\text{norm}}$: 
    $\lambda_{(1)} \le \cdots \le \lambda_{(n)}$\;
    \eIf{$K_{\text{true}}$ is not null}{
      Set $K(\alpha) \leftarrow K_{\text{true}}$\;
    }{
      Set $K(\alpha) \leftarrow \argmax_{k \in 
        \{2, \ldots, n - 1\}} \left(\lambda_{(k+1)} - \lambda_{(k)} 
      \right)$\;
    }
    Extract community labels $\bm{z}(\alpha)$ by applying the 
    $k$-means algorithm to the eigenvectors corresponding to
    $\lambda_{(1)}, \ldots, \lambda_{(K(\alpha))}$\;
    Compute the silhouette score $\text{SilS}(\alpha)$ based on 
    $\bm{C}$ and $\bm{z}(\alpha)$ using distance
    $d(i,j) = 1 / (C_{ij} + \varepsilon)$\;
}
Set $\alpha_{\text{opt}} \leftarrow 
\argmax_{\alpha \in \mathbb{A}} \text{SilS}(\alpha)$,
$\bm{z} \leftarrow \bm{z}(\alpha_{\text{opt}})$, and
$K_{\text{opt}} \leftarrow K(\alpha_{\text{opt}})$\;
\Return{$\bm{z}$, $\alpha_{\text{opt}}$, and $K_{\text{opt}}$.}
\end{algorithm}

The pseudocode for $\text{C}^4$ is provided in 
Algorithm~\ref{alg:c4}. An open-source and user-friendly 
\proglang{R} package, \texttt{c4}, is available on GitHub. The 
package provides functions for performing spectral clustering, 
including the proposed $\text{C}^4$ and the covariate-assisted 
spectral clustering (CASC) method 
by~\citet{binkiewicz2017covariate}, along with additional utilities 
for preprocessing and evaluation. Although several CASC 
implementations
are publicly available, the original \texttt{rCASC} codes from
\citet{binkiewicz2017covariate} did not run
properly in our experiment, mainly due to incompatibility with updates
of several supporting \proglang{R} package. In addition, the
\texttt{CASC()} function in the
\texttt{CASCORE} package~\citep{pkg:CASCORE} applies only when the
covariate dimension exceeds the number of
clusters, which limits its applicability. Hence, we developed our 
own implementation of CASC in package \texttt{c4}, with the full 
workflow illustrated in a vignette, from data generation and 
preprocessing to clustering. All random seeds used in our 
simulations are properly documented to ensure reproducibility.

\section{Simulations}
\label{sec:sim}

Simulation study evaluates the performance of the proposed
$\text{C}^4$ method under controlled network topology and covariate
signal conditions. The study consists of two parts. The first 
examines scenarios where the number of clusters~$K$ is known, while 
the second considers a more realistic setting where $K$ is unknown. 
In both parts, $\text{C}^4$ is compared with two fixed baselines: 
(1) structure-only clustering, which applies
spectral clustering to the weighted adjacency matrix~$\bm{W}$ 
(corresponding to $\alpha = 0$); and (2) covariate-only clustering,
which applies spectral clustering to the covariate similarity
matrix~$\bm{S}$ (corresponding to $\alpha = 1$). Unlike these
baselines, $\text{C}^4$ adaptively selects~$\alpha$ from the tuning
grid $\mathbb{A} = \{0, 0.1, 0.2, \ldots, 1\}$ to optimize clustering
performance. When~$K$ is known, CASC is included for
comparison~\citep{binkiewicz2017covariate}. All simulation codes and 
random seeds are provided in the online supplements to ensure 
reproducibility.

\subsection{Data Generation}
\label{sec:datagen}

The simulation design systematically varies network size, cluster 
number, and covariate and structural signal strengths to evaluate 
$\text{C}^4$ under conditions of increasing difficulty. Networks of 
size $n \in \{400, 800\}$ were generated with $K \in \{4, 8\}$ 
communities. The strength of covariate signal was jointly determined 
by the distances among cluster means and the noise 
parameter~$\sigma$. Specifically, each node was associated with a 
three-dimensional covariate vector ($p = 3$) drawn from a 
multivariate normal distribution $\mathrm{MVN}(\bm{\mu}_k, \sigma^2 
\bm{I})$ if node~$i$ belongs to cluster $k \in \{1, 2, \ldots, K\}$, 
where $\bm{\mu}_k \in \mathbb{R}^3$ denotes the cluster mean and
$\sigma$ controls within-cluster variability.

The covariate structure controlled overlap among communities
through the configuration of cluster centers and the noise
parameter~$\sigma$. When $K = 4$, the cluster centers~$\bm{\mu}_k$
corresponded to the four vertices of a regular tetrahedron centered at
the origin with fixed edge length~$10$. When $K = 8$, they were 
positioned at the eight vertices of a cube centered at the origin 
with the same edge length. The degree of overlap, and thus the 
strength of the covariate signal, was determined jointly by the 
distances between cluster means and the value of~$\sigma$. We 
considered~$\sigma \in \{2, 3\}$, representing respectively strong 
and weak separation. Pairwise dissimilarity between nodes was 
measured by the Euclidean distance, and the similarity 
matrix~$\bm{S}$ was defined as $s_{ij} = 1 / \Vert  \bmX_i - \bmX_j 
\Vert_2$ for $i \neq j$ and $s_{ii} = 0$. This design provided 
continuous covariates with tunable signal strength for evaluating 
information fusion in clustering.

On the other hand, the strength of the structural signal was 
controlled by edge probabilities and weights. Specifically, network 
topology was generated by fixing the within-community edge
probability at $b_{\text{win}} = 0.6$ and varying the
between-community probability $b_{\text{btw}} \in \{0.3, 0.4, 0.5\}$.
Both unweighted and weighted networks were examined. In the weighted
case, each edge weight~$w_{ij}$ was independently drawn from
$\mathrm{Gamma}(2, \theta)$ with $\theta_{\text{btw}} = 1$ for
between-community edges and
$\theta_{\text{win}} \in \{1,\,1.25,\,1.5\}$ for within-community
edges. Increasing $b_{\text{btw}}$ or decreasing $\theta_{\text{win}}$
weakened the structural contrast between communities, making the
clustering problem more difficult. In contrast, larger network size
$n$ improved estimation stability and enhanced separation. Each
combination of $(n, K, \sigma, b_{\text{btw}})$ and weighting scheme,
either unweighted or weighted with a given $\theta_{\text{win}}$, was
replicated 100 times to ensure reliable comparisons across methods.

\subsection{Results}

We separately report the results from settings with known $K$ and
unknown~$K$.

\subsubsection{Known K}
\label{sec:simu-known}

\begin{figure}[tbp]
  \centering
  (a) Network size $n = 400$.\\
  \includegraphics[width=\textwidth]{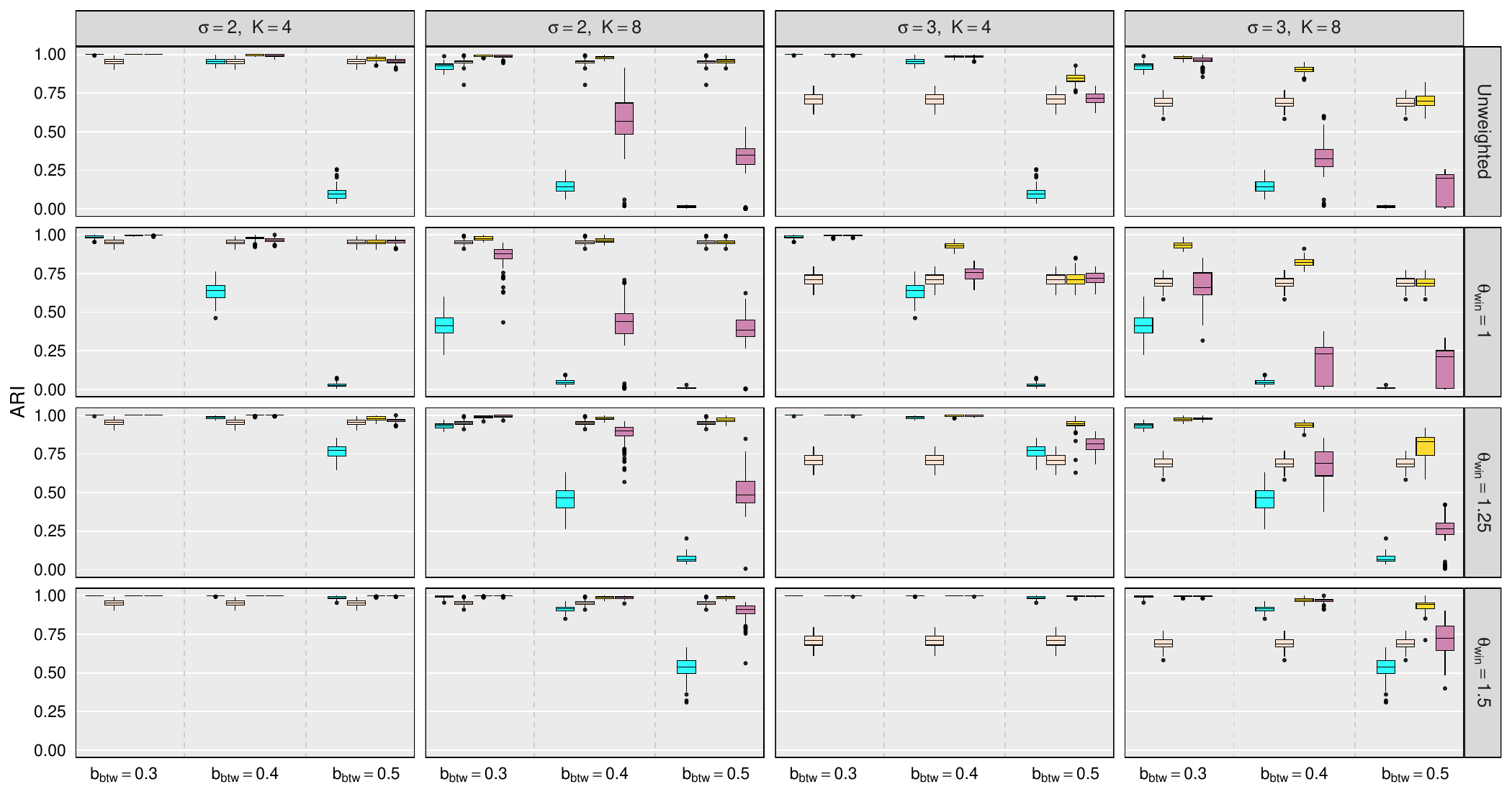}
  \par\vspace{2em}
  (b) Network size $n = 800$.\\
  \includegraphics[width=\textwidth]{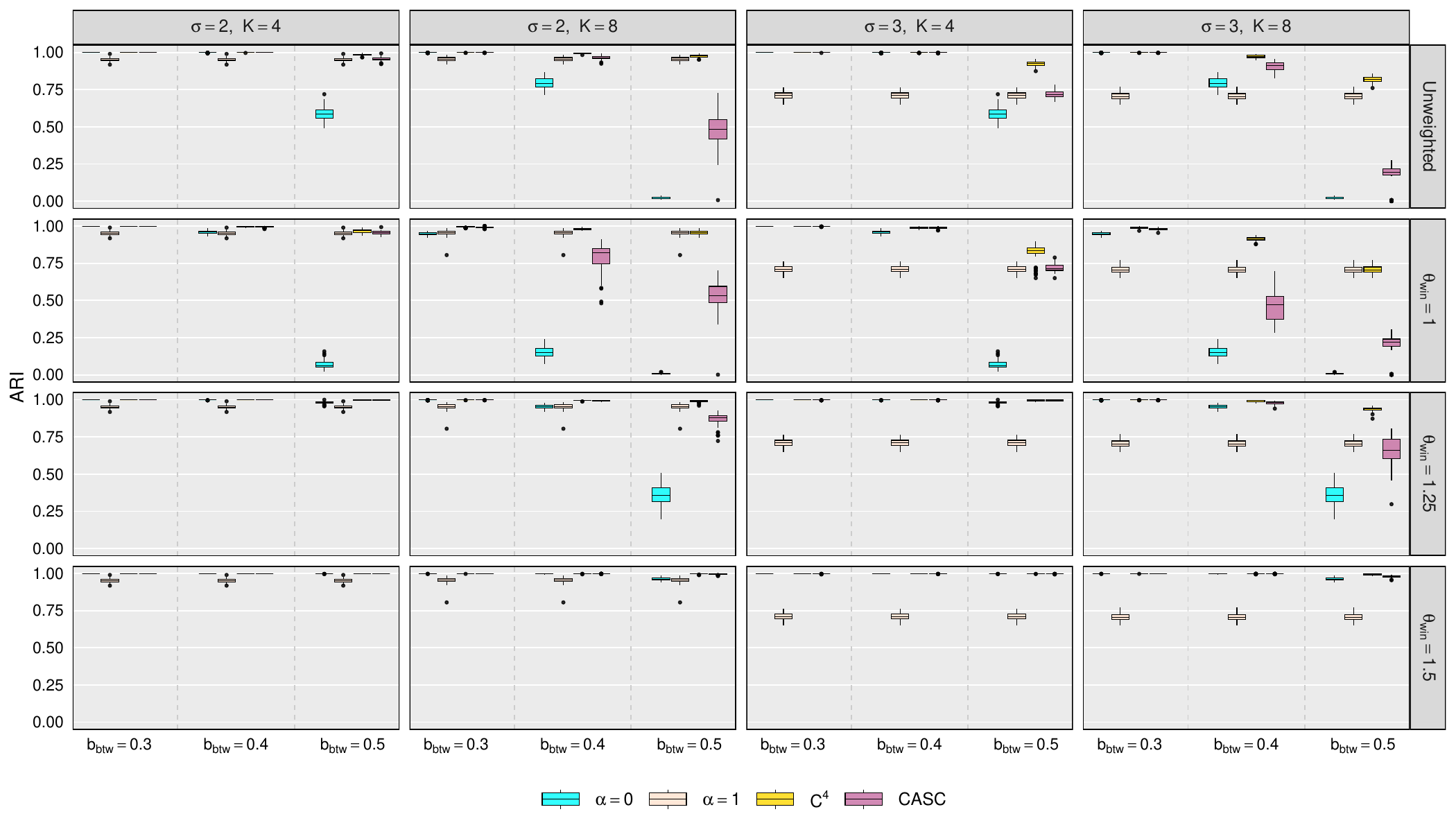}
  \caption{Side-by-side ARI boxplots for known~$K \in 
  \{4, 8\}$ with $n = 400$ (upper) and $n = 800$ (lower).}
  \label{fig:known_box_n}
\end{figure}

The experiments with known~$K$ evaluated how accurately and
consistently $\text{C}^4$ recovered community structures when the true
number of clusters was provided. Clustering accuracy was measured by
the Adjusted Rand Index~\citep[ARI,][]{rand1971objective}, which
quantifies agreement between estimated and true community
assignments. ARI is unaffected by label switching and remains stable
under unbalanced group sizes. Figure~\ref{fig:known_box_n} displays
boxplots of the ARI from 100 replicates under each simulation setting
for different methods.

Across all settings, weakening the structural signal lowers accuracy
for the baselines, while $\text{C}^4$ remains stable. In unweighted
networks, holding $\sigma$, $n$, $K$ and $b_\text{win} = 0.6$ fixed, 
increasing $b_{\text{btw}}$ from~0.3 to~0.5 reduces within- and 
between-community contrast and drives the ARI for $\alpha=0$ sharply 
downward in every panel of the top row of 
Figure~\ref{fig:known_box_n}. CASC tracks $\text{C}^4$ when the
structure is strong (e.g., $b_{\text{btw}}=0.3$) but deteriorates more
quickly as $b_{\text{btw}}$ grows, with lower medians and greater
dispersion. The same pattern appears in weighted networks as
$\theta_{\text{win}}$ decreases. For example, in
Figure~\ref{fig:known_box_n} with $\sigma=2$, $n=400$, and $K=8$,
$\alpha=0$ continues to fall as $\theta_{\text{win}}$ moves 
from~$1.5$ to~$1$, CASC declines noticeably, and $\text{C}^4$ 
changes little with small variance. These comparisons imply that CASC 
is more sensitive to structural weakening, whereas $\text{C}^4$ 
maintains accuracy and stability.

When the covariate signal weakens, $\alpha=1$ drops, but $\text{C}^4$
remains stable and often outperforms. Comparing 
Figure~\ref{fig:known_box_n} at $\sigma=2$ versus $\sigma=3$ with
$n=400$ and $K=8$, both $\text{C}^4$ and CASC achieve high ARI when 
the structure is strong (e.g., $\theta_{\text{win}}=1.5$ with
$b_{\text{btw}}\in\{0.3,0.4\}$), indicating limited sensitivity to
covariate noise under dominant structural topology. As structural 
information becomes limited, increasing $\sigma$ depresses CASC more 
than $\text{C}^4$: medians shift downward and variability grows 
across all three $b_{\text{btw}}$ levels, while $\text{C}^4$ retains 
higher medians and smaller dispersion. These patterns show that when 
covariate signal is weaker and structure is not decisive, 
$\text{C}^4$ better balances the two sources and preserves accuracy.

Larger networks and fewer communities aid all methods, but $\text{C}^4$
continues to lead in median ARI and stability. Gains are evident as $n$
increases or $K$ decreases across panels, reflecting more stable
eigenvectors and easier separation. Notably, under the most extreme
setting ($b_{\text{btw}}=0.5$, $\theta_{\text{win}}=1$, $K=8$), the
structure provides little usable information; $\text{C}^4$ then selects
$\alpha=1$ and relies on covariates, yielding reasonable clustering
despite negligible topology. While such cases are rare in practice,
they illustrate the adaptive mechanism: $\text{C}^4$ shifts weight 
away from uninformative structure to maintain performance.

\subsubsection{Unknown K}
\label{sec:simu-unkown}

The next experiment considered the more realistic setting in which the
number of communities~$K$ was unknown. For each method, $K$ was
determined using the eigengap heuristic described in
Section~\ref{sec:c4}, and clustering accuracy was again evaluated by 
ARI.

The accuracy of $K$~estimation was examined first. 
Table~\ref{tab:K_dist} presents a representative  example with 
moderate topology structure: $n = 400$, $b_{\text{btw}} = 0.4$, and 
$K = 8$, using edge weights from a Gamma distribution with scale 
$\theta_{\text{win}} = 1.25$, a configuration typical of many 
weighted networks where within-community edges have larger average 
weights. Results are reported for two levels of the
signal-to-noise ratio controlled by $\sigma \in \{2, 3\}$.

\begin{table}[tbh]
    \centering
    \caption{A summary of the selected community number $K$ for $n = 400$, 
    $b_{\text{btw}} = 0.4$,~$\theta_{\text{win}}=1.25$, and true $K = 
    8$: Sensitivity to the signal-to-noise ratio $\sigma \in \{2, 
    3\}$.}
    \setlength{\tabcolsep}{7pt}
    \begin{tabular}{lcccccccccccccccc}
    \toprule
     & \multicolumn{8}{c}{\text{$\sigma = 2$}} & \multicolumn{8}{c}
     {\text{$\sigma = 3$}} \\ 
    \cmidrule(lr){2-9} \cmidrule(lr){10-17}
    \text{Method} & 2 & 3 & 4 & 5 & 6 & 7 & 8 & $\ge9$
                  & 2 & 3 & 4 & 5 & 6 & 7 & 8 & $\ge9$ \\
    \midrule
    $\alpha = 0$ & \textcolor{blue}{34} & 20 & 17 & 8 & 11 & 7 & 1 & 2 
                  & \textcolor{blue}{34} & 20 & 17 & 8 & 11 & 7 & 1 & 2 \\
    $\alpha = 1$ & 2 & 0 & \textcolor{blue}{59} & 1 & 0 & 0 & 34 & 4
                  & 2 & 0 & \textcolor{blue}{95} & 3 & 0 & 0 & 0 & 0 \\
    $\text{C}^4$          & 0 & 0 & 0 & 0 & 0 & 0 & \textcolor{blue}{98} & 2
                  & 1 & 0 & \textcolor{blue}{56} & 3 & 1 & 7 & 30 & 2 \\
    \bottomrule
    \end{tabular}
    \label{tab:K_dist}
\end{table}

The baseline method $\alpha = 0$ underestimates the number of 
clusters. This poor performance reflects the difficulty of 
extracting sufficient community information under moderate network 
structures. When the covariate signal is strong ($\sigma = 2$), 
$\alpha = 1$ performs reasonably well, correctly identifying $K = 8$ 
in about one-third of cases, but it also tends to underestimate $K$. 
By integrating both structure and covariate information, $\text{C}^4$ 
achieves near-perfect accuracy, significantly outperforming either 
baseline. This demonstrates that when both structure and covariate 
are informative, $\text{C}^4$ can leverage their complementary 
strengths to achieve higher accurate $K$ estimation. As the 
covariate signal weakens to $\sigma=3$, $\alpha = 1$ exhibits a 
dramatic decline, with correct estimated $K$ falling to zero. 
However, $\text{C}^4$ maintains reasonable performance, correctly 
identifying $K=8$ in $30\%$ cases. Crucially, the adaptive mechanism 
of $\text{C}^4$ allows it to compensate for the 
weakened covariate signal by increasing the weight on structural 
information. While neither structure alone ($\alpha = 0$) nor 
covariate alone ($\alpha = 1$) provides accurate $K$ estimation in 
this challenging scenario, integration of both sources under 
$\text{C}^4$ yields more reliable results. The patterns observed in 
Table~\ref{tab:K_dist} are consistent among other parameter 
combinations (see Appendix~\ref{sec:app} for the distributions of $K$ 
selection), highlighting that the advantage of $\text{C}^4$ in $K$ 
selection remains stable across different structural and covariate 
signal strengths.

\begin{figure}[tbp]
  \centering
  (a) Network size $n = 400$.\\
  \includegraphics[width=\textwidth]{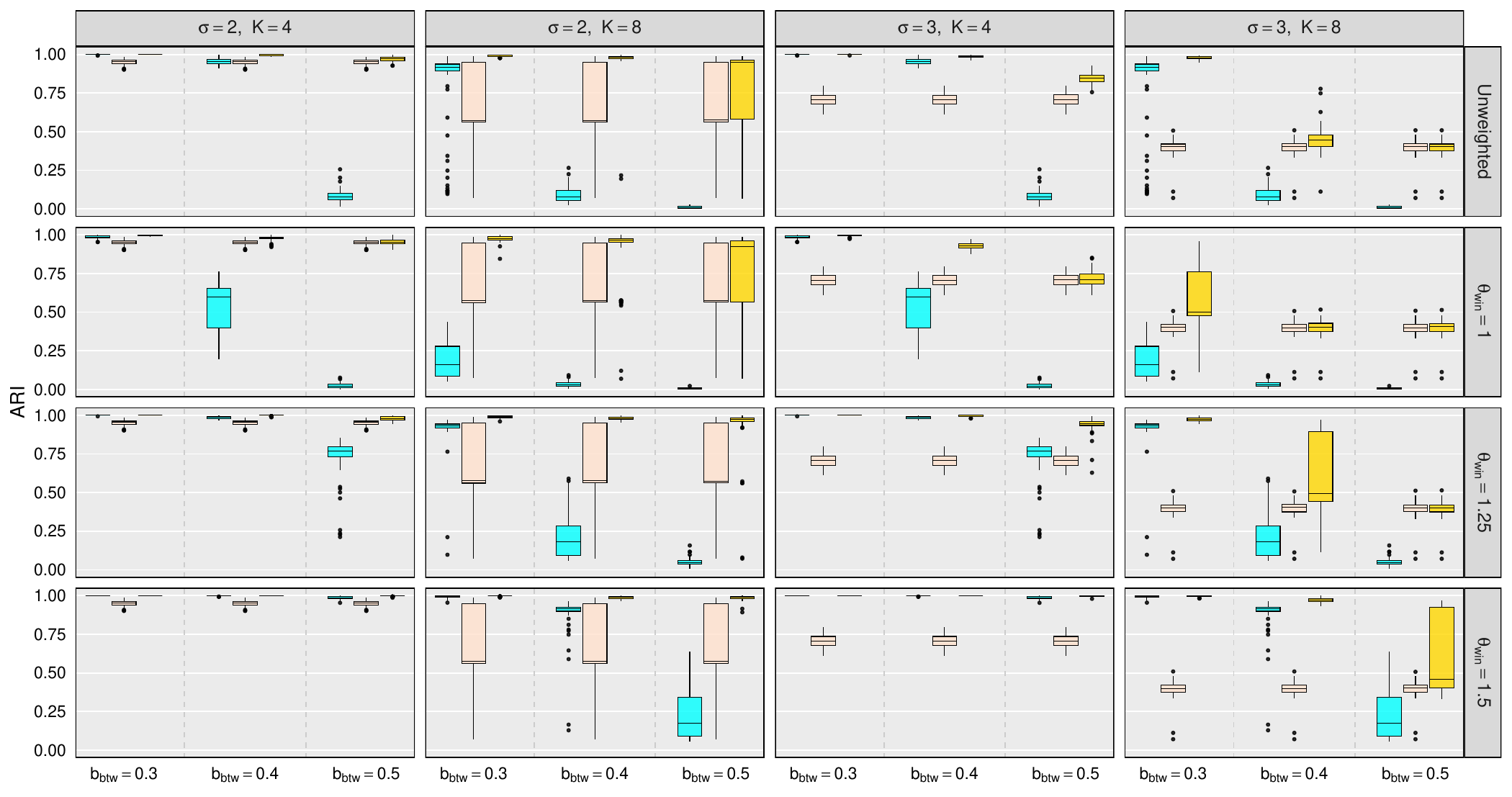}
  \par\vspace{2em}
  (b) Network size $n = 800$.\\
  \includegraphics[width=\textwidth]{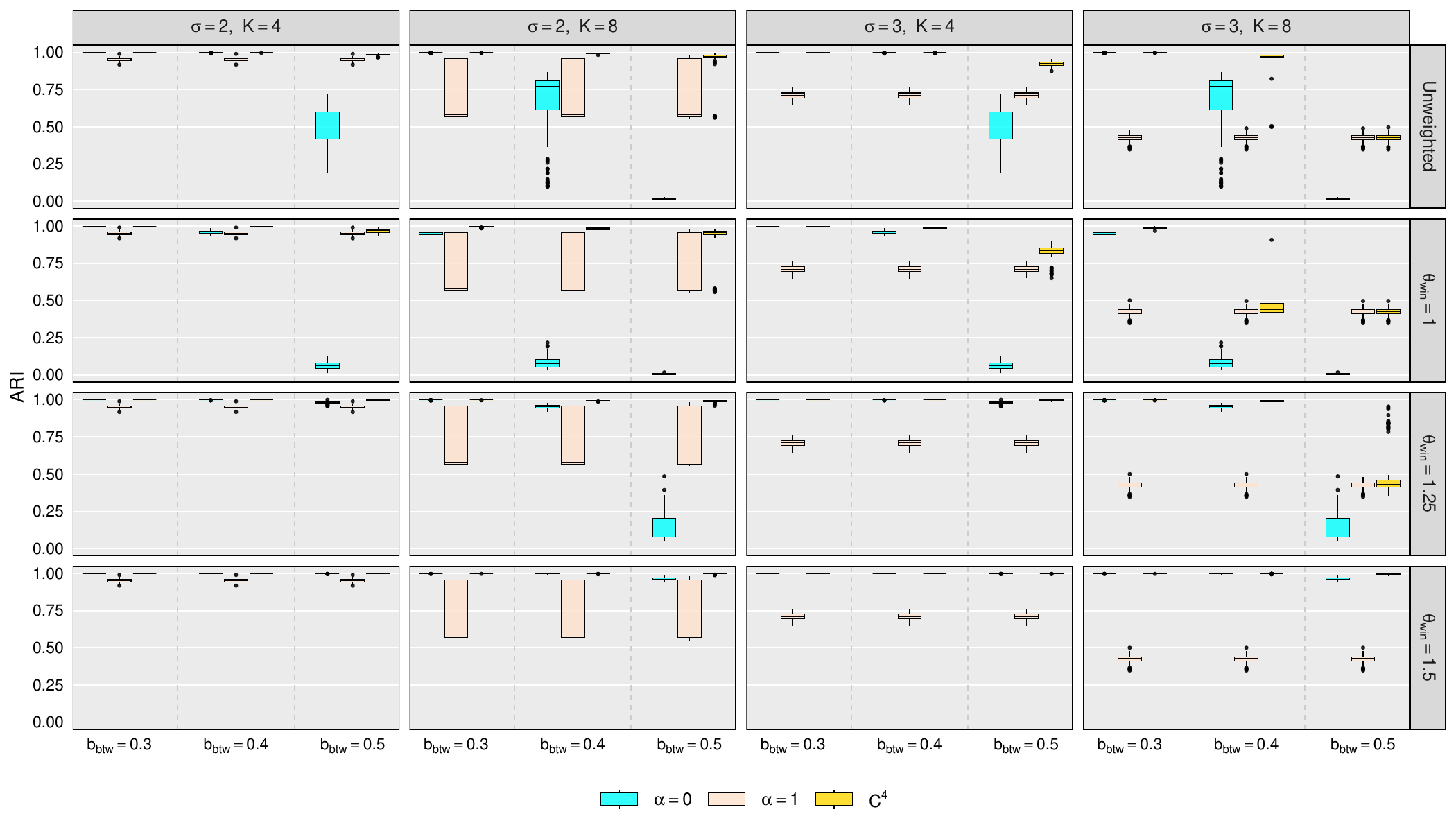}
  \caption{Side-by-side ARI boxplots for unknown~$K \in 
  \{4, 8\}$ with $n = 400$ (upper) and $n = 800$ (lower).}
  \label{fig:unknown_box_n}
\end{figure}

Clustering accuracy with unknown~$K$ remains highest and most stable 
for $\text{C}^4$ across all simulation settings. Boxplots in
Figure~\ref{fig:unknown_box_n} show that, although every method 
suffers some loss of accuracy relative to the known $K$ case
(Figure~\ref{fig:known_box_n}), $\text{C}^4$ consistently attains the
highest or tied-for-highest median ARI with smaller dispersion. When
both structural and covariate information are strong (e.g.,
$\sigma = 2$, $K = 4$, $n = 400$, $b_{\text{btw}} = 0.3$), 
$\text{C}^4$ effectively fuses the two sources and matches or 
surpasses the best baseline performance. When only one source is 
reliable, either strong topology but weak covariates ($\sigma = 3$) 
or strong covariates but weak topology ($b_{\text{btw}} = 0.5$), 
$\text{C}^4$ automatically assigns greater weight to the informative 
component, maintaining high and stable ARI. Even when both signals 
are weak, $\text{C}^4$ produces superior accuracy to either
baseline alone. Patterns remain consistent across weighted networks 
and different values of~$n$ and~$K$. In extreme configurations 
characterized by weak structure and noisy covariates
($b_{\text{btw}} = 0.5$, $\sigma = 3$, $K = 8$),
$\text{C}^4$ tends to select $\alpha = 1$, effectively reducing to
covariate-only clustering; details of these cases and the resulting
$K$~distributions are provided in Appendix~\ref{sec:app}.

\section{Application to Airline Reachability Network}
\label{sec:real_data}

We applied our method to an Airline Reachability Network (ARN)
dataset~\citep{frey2007clustering}. The dataset 
captures the reachability relationships among 456 major cities in 
the United States (including Alaska and Hawaii) and 
Canada, reflecting how easily each city can be accessed by others in 
terms of estimated commercial airline travel times. Each node 
represents a city, and a directed edge from city $i$ to 
city~$j$ is established if the estimated airline travel time from $i$ 
to $j$, including expected stopover delays, does not exceed 48 hours.
The travel times in ARN range from 10 minutes for short regional
flights to nearly 48 hours for the longest multi-leg itineraries.
Because airline routes are not always reciprocal and prevailing 
headwinds cause directional differences in flight durations, the ARN 
exhibits both structural and weight asymmetries: some city pairs are 
connected only in one direction due to route availability, and even 
for bidirectional pairs, the edge weights from $i$ to $j$ and from 
$j$ to $i$ differ because of flight-time asymmetry. Among the 71,959 
directed edges, 34,012 city pairs are bidirectional and 3,935 are 
unidirectional. Among the bidirectional edges, 98\% have different weights.

We construct an undirected, weighted network by including only city
pairs that are mutually connected via directed edges. The edge weight 
is computed as the inverse of the mean bidirectional travel time 
between nodes $i$ and $j$ so that shorter travel times correspond to 
stronger connections. The resulting undirected ARN exhibits 
substantial degree heterogeneity: a small number of hub cities 
connect to hundreds of cities, whereas most smaller regional cities 
have only a few connections. Such structural characteristics make 
the ARN a challenging but informative setting for evaluating our 
$\text{C}^4$ method.

We incorporate a single node-level covariate for each city: the logarithm
of the surrounding metropolitan population~\citep{benson2016higher}.
Population size captures the scale of travel demand and the relative
importance of a city within the airline mobility system, making it a
meaningful factor for characterizing node-level heterogeneity. For each
city pair, we define similarity as the mean of their log-transformed
populations. Using the mean emphasizes the overall market size accessible
through a connection and captures the shared demand potential between two
cities. This generates a symmetric similarity matrix that serves as 
the
covariate-based input to our method.

\begin{figure}[tbp]
    \centering
    \includegraphics[width=0.95\textwidth]{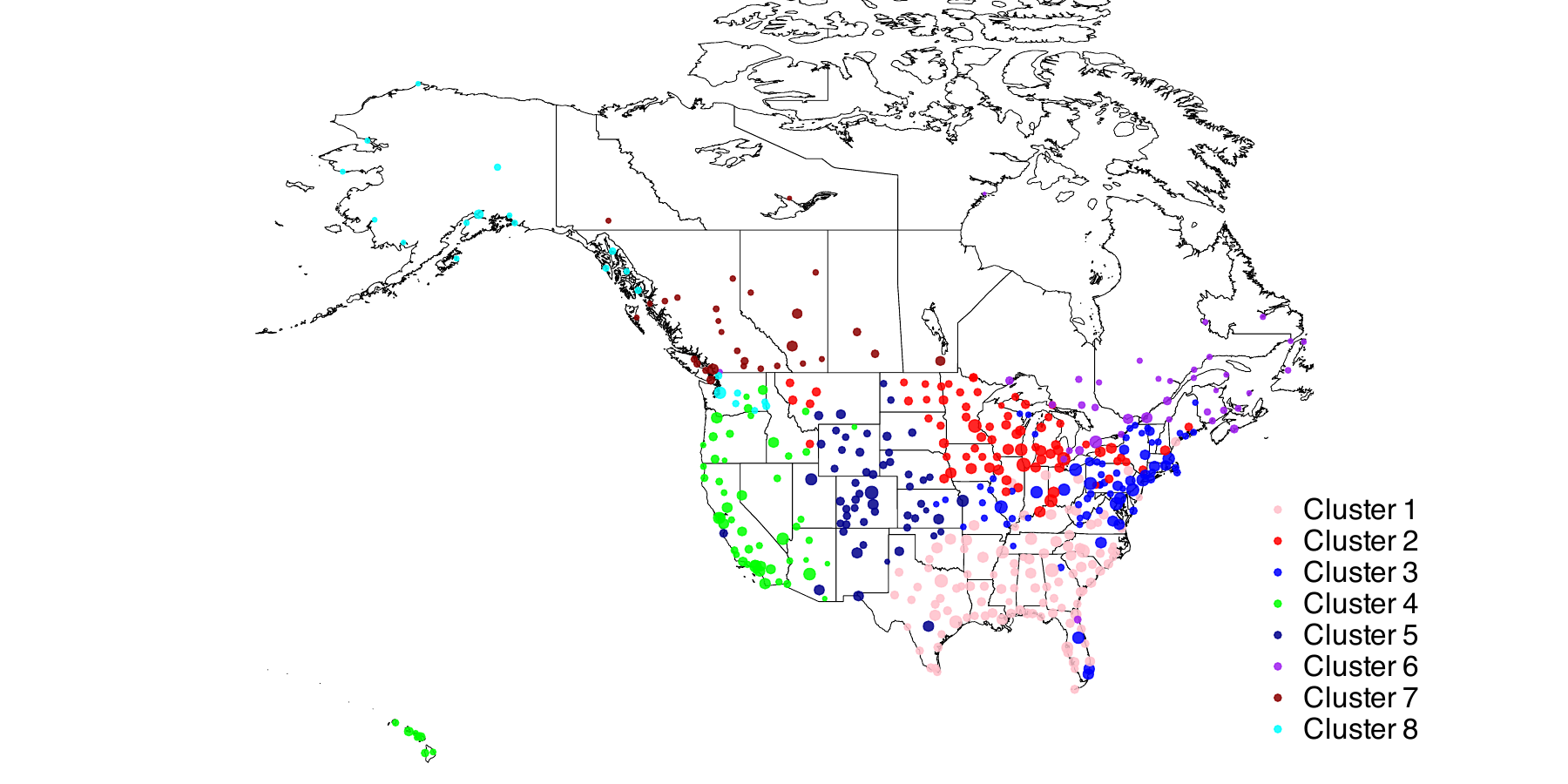}
    \caption{Clustering result based on application of $\text{C}^4$ to the 
    ARN.}
    \label{fig:ARN_plot}
\end{figure}

\begin{table}[tbp]
\centering
\caption{Summary of community properties of airports across the U.S. 
and Canada, including community size, largest city (in terms of 
node strength), and metropolitan population median.}
\label{tab:ARN_table}
\setlength{\tabcolsep}{12pt}
\begin{tabular}{lrrr}
\toprule
 & Size & Largest City & Population Median \\
\midrule
Cluster 1 & 106 & Atlanta, GA       & 357{,}728 \\
Cluster 2 &  75 & Chicago, IL       & 187{,}612 \\
Cluster 3 &  74 & Washington, DC    & 133{,}310 \\
Cluster 4 &  59 & Phoenix, AZ       & 211{,}888 \\
Cluster 5 &  54 & Denver, CO        &  46{,}388 \\
Cluster 6 &  37 & Toronto, ON       &  53{,}100 \\
Cluster 7 &  29 & Vancouver, BC     &  46{,}850 \\
Cluster 8 &  22 & Seattle, WA       &  13{,}732 \\
\bottomrule
\end{tabular}
\end{table}

\begin{table}[tbp]
	\centering
	\caption{Within- and between-community densities (in 
		percentage) based on the clustering result.}
	\label{tab:density}
	\setlength{\tabcolsep}{12pt}
	\begin{tabular}{lcccccccc}
		\toprule
		& C-1 & C-2 & C-3 & C-4 & C-5 & C-6 & C-7 & C-8 \\
		\midrule
		C-1 & $67$ & $47$ & $39$ & $35$ & $29$ & $14$ & $11$ & $11$ \\
		C-2 &  & $68$ & $36$ & $33$ & $33$ & $16$ & $16$ & $14$ \\
		C-3 &  &  & $47$ & $29$ & $24$ & $22$ & $12$ & $13$ \\
		C-4 &  &  &  & $54$ & $45$ & $12$ & $22$ & $36$ \\
		C-5 &  &  &  &  & $71$ & $8$ & $15$ & $15$ \\
		C-6 &  &  &  &  &  & $45$ & $28$ & $5$ \\
		C-7 &  &  &  &  &  &  & $68$ & $14$ \\
		C-8 &  &  &  &  &  &  &  & $71$ \\
		\bottomrule
	\end{tabular}
\end{table}

Applying the $\text{C}^4$ method to the ARN leads to the optimal 
tuning parameter $\alpha = 0.4$ and selects $K=8$ based on the 
eigengap heuristic. Figure~\ref{fig:ARN_plot} depicts the community 
structure, with node size proportional to strength (the sum of 
incident edge weights), revealing geographically coherent clusters. 
For clarity, edges between nodes are omitted in 
Figure~\ref{fig:ARN_plot}. Table~\ref{tab:ARN_table} reports each 
cluster’s size, largest city (by node strength), and median 
metropolitan population, while Table~\ref{tab:density} presents both 
within- and between-cluster densities, defined as the proportion of 
observed edges among all possible node pairs, with clusters denoted 
by C-1 to C-8. The results highlight four key 
characteristics driven by population scale and connection intensity: 
(1) Clusters~1--2 contain large metropolitan areas such as Atlanta 
and Chicago, combining high population with strong internal ties 
that form national travel corridors; (2) Clusters~3--4 include major 
cities such as Washington~DC, New~York, Phoenix, and Los~Angeles, 
representing equally large markets but lower densities due to wide 
geographic span; (3) Clusters~5, 7, and 8 comprise metropolitan 
cities such as Denver, Vancouver, and Seattle that (though not as 
large as New York or Los Angeles) exhibit high internal cohesion.; 
and (4) Cluster~6, with Toronto among its
largest members, spans distant Canadian provinces and exhibits the
weakest internal linkage. Overall, the $\text{C}^4$ results show that
population scale and connectivity intensity are not aligned and that 
the method effectively disentangles these two structural dimensions 
of the airline network.

Clusters~1--4 contain populous U.S.\ regions that anchor the national
air-transportation system but differ markedly in their internal
connectivity. Cluster~1, centered on Atlanta, the largest city in the
group, extends through the Southeast and into the eastern Midwest, 
while Cluster~2, led by Chicago, links central U.S.\ markets with 
the East Coast. In contrast, Cluster~3 comprises several global 
metropolitan centers including Washington~DC, New~York, 
Philadelphia, and Boston, yet extends southward to tourism-linked 
cities such as Orlando and Miami. Cluster~4, encompassing Phoenix, 
Los~Angeles, Las~Vegas, San~Francisco, and tourist destinations in 
Hawaii, comprises major metropolitan areas in the western region and 
exhibits a hub-dominated structure.

Clusters~5--8 represent smaller but more cohesive regional systems,
largely situated in the Mountain West and adjacent western regions of
the United States and Canada. Clusters~5, 7, and~8 share a common
pattern of compact geography and strong internal ties despite modest
population size. Cluster~5, centered on Denver, connects numerous 
Rocky Mountain cities into a dense network. Cluster~7, dominated by 
Vancouver, forms a tightly connected western Canadian region, and 
Cluster~8 links Seattle with airports in 
Alaska and the Pacific Northwest, making it the smallest yet most 
internally connected community. In contrast, Cluster~6, with Toronto 
as its largest city, spans central Canadian provinces, reflecting 
long distances between major population centers therein. These 
patterns demonstrate that compact geography, rather than market 
size, drives cohesion, a distinction the $\text{C}^4$ method 
captures effectively within the airline network.

\section{Discussion}
\label{sec:disc}

This study proposes a novel clustering framework, $\text{C}^4$, which 
adaptively integrates network topology and node-level covariate 
information through a data-driven weighting mechanism. The method 
unifies structure- and covariate-based community detection within a 
single framework that allows both the tuning parameter $\alpha$ and 
the number of clusters $K$ to be automatically selected or flexibly 
specified by the user. Because the covariate similarity matrix 
$\bm{S}$ is normalized to match the scale of the adjacency matrix 
$\bm{W}$, the tuning parameter~$\alpha$ gains a clear and 
interpretable meaning, in contrast to the CASC 
algorithm~\citep{binkiewicz2017covariate}, where such interpretation 
is not directly available due to the mismatch of scales. 
Simulation results show that $\text{C}^4$ consistently achieves 
accurate and stable recovery of community structures across diverse 
configurations. The algorithm remains computationally efficient, with 
a complexity comparable to standard spectral clustering, and scales 
well to networks with thousands of nodes, making it suitable for 
practical applications.

Despite these strengths, several limitations and extensions merit 
future investigation. First, when the covariate information consists 
of a single categorical variable, the silhouette score may become 
degenerate, achieving its maximum value of~$1$ when the clustering is 
trivially determined by the covariate~($\alpha = 1$). This degeneracy 
prevents the adaptive selection of~$\alpha$, as the silhouette 
criterion becomes uninformative under perfectly separable partitions. 
Alternative selection strategies or penalized objectives that 
down-weight trivially separable configurations could mitigate this 
issue. Second, the current implementation relies on a fixed grid 
search over~$\alpha$ and a chosen covariate similarity measure; finer
grids or alternative distance metrics may influence performance. 
Future work could formalize stability diagnostics or develop 
data-driven procedures for choosing both the similarity function and 
tuning grid. Third, while $\text{C}^4$ effectively handles weighted 
networks, further research could examine how heterogeneous degree 
distributions or extreme weight variability affect clustering results 
and whether normalization or degree-correction improves robustness. 
Finally, establishing theoretical guarantees for community recovery 
and extending the framework to dynamic or multilayer networks would 
enhance its general applicability.

\appendix
\section{Appendix}
\label{sec:app}

\begin{figure}[tbp]
  \centering
  (a) Network size $n = 400$.\\
  \includegraphics[width=\textwidth]{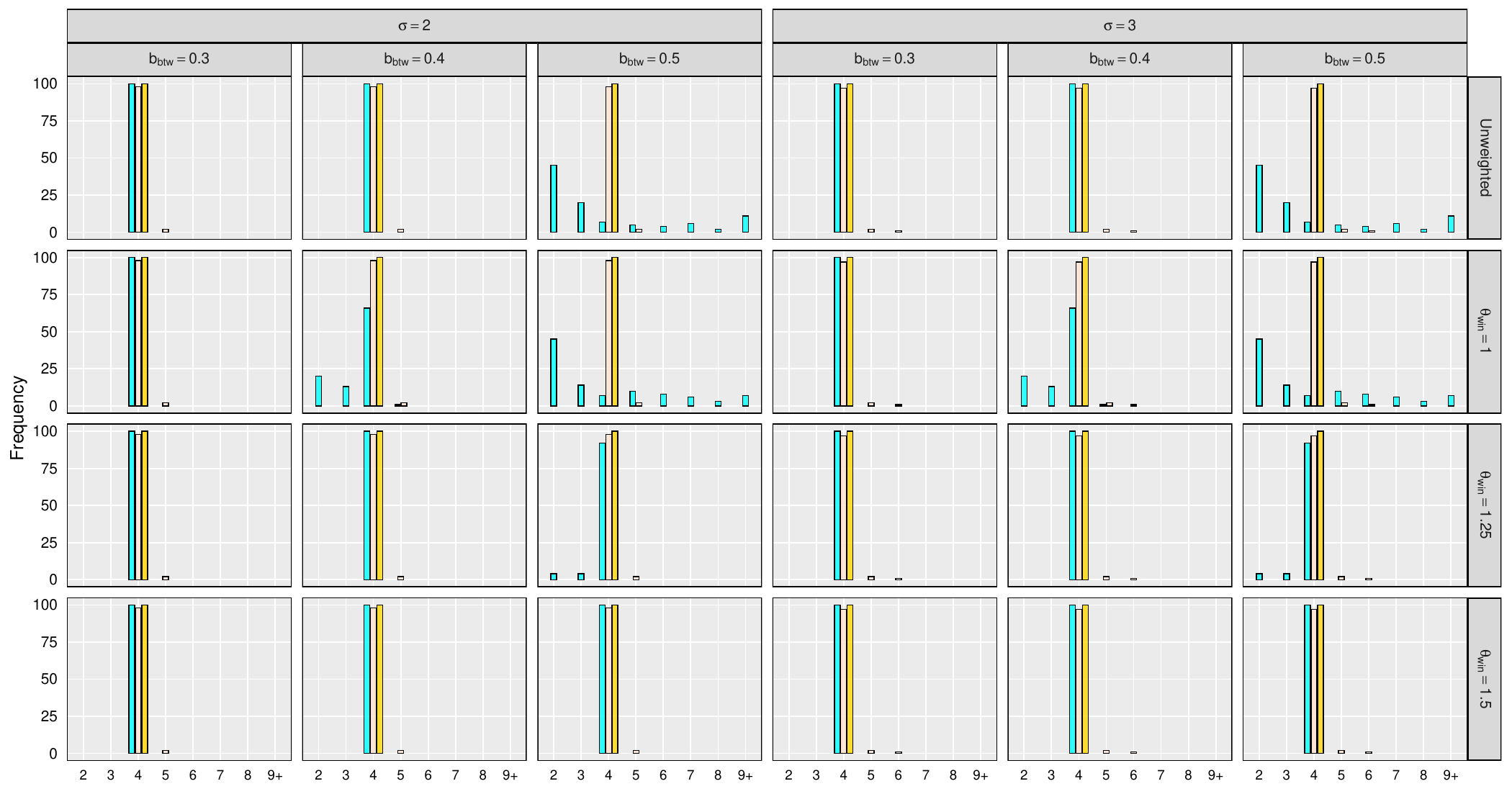}
  \par\vspace{2em}
  (b) Network size $n = 800$.\\
  \includegraphics[width=\textwidth]{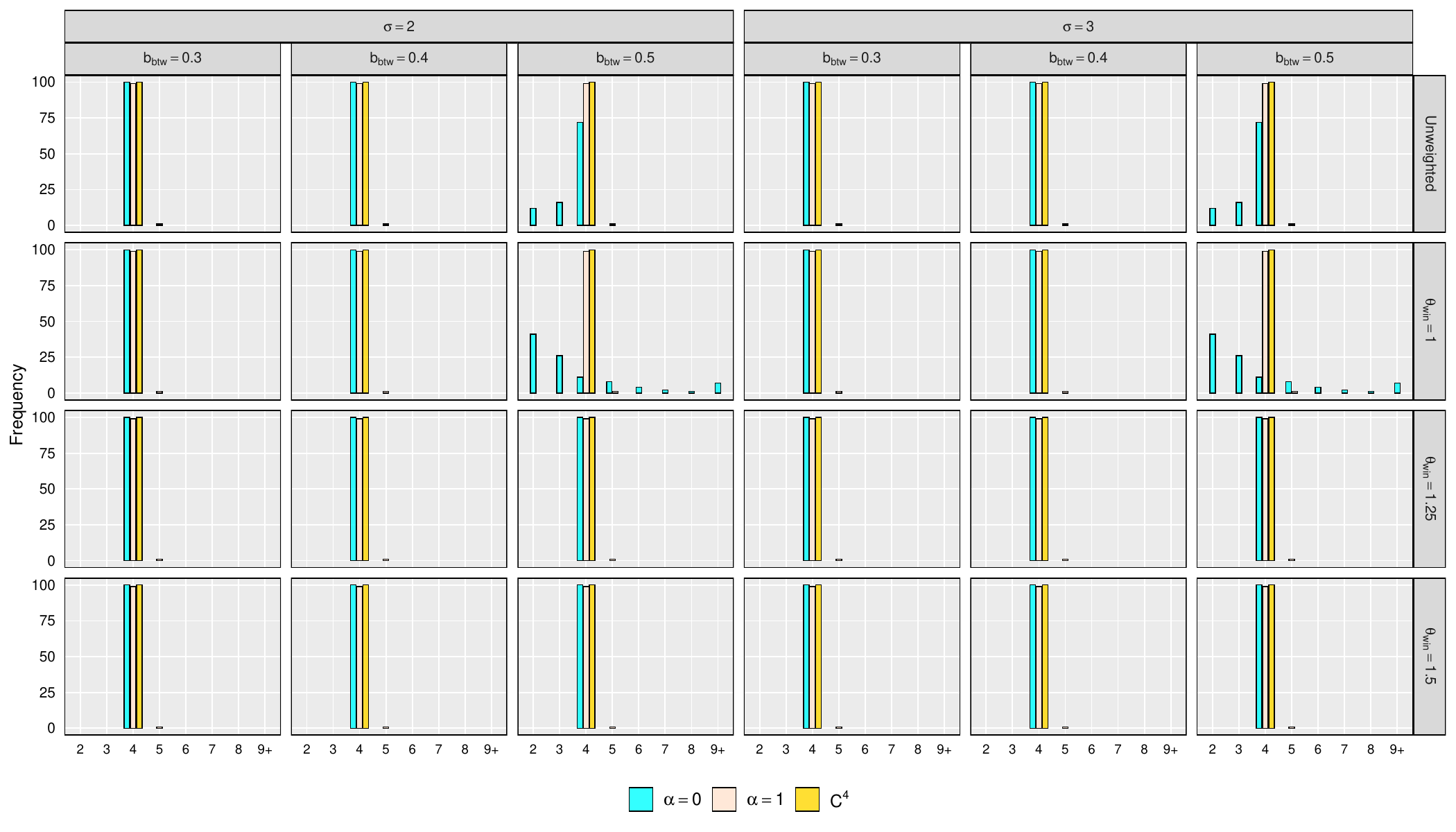}
  \caption{Distribution of the $K$ selection  under 
  $K_{\text{true}}=4$ among all simulation configurations with 
  $n=400$ (upper) and $n=800$ (lower).}
  \label{fig:K4_dist}
\end{figure}

\begin{figure}[tbp]
  \centering
  (a) Network size $n = 400$.\\
  \includegraphics[width=\textwidth]{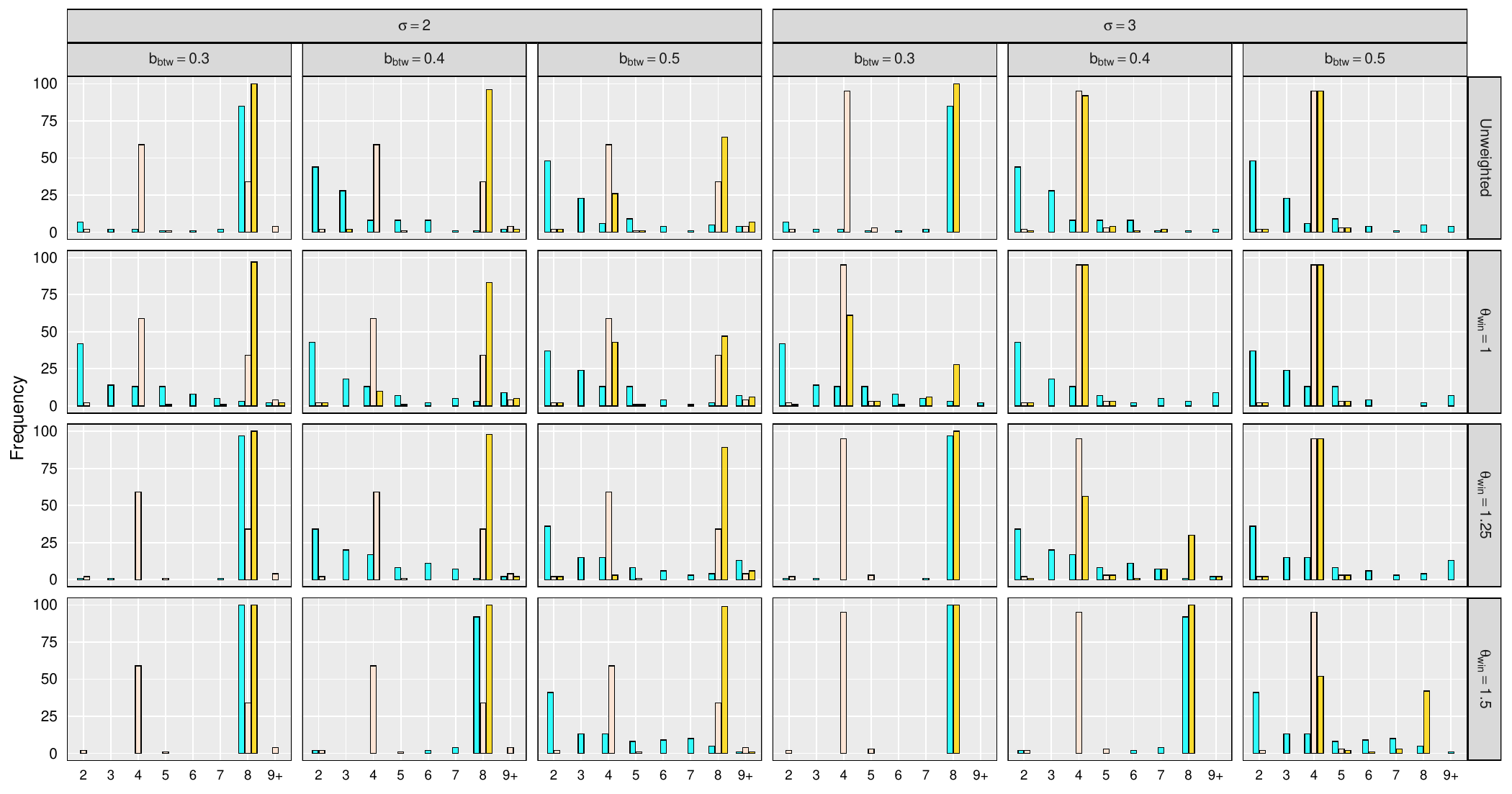}
  \par\vspace{2em}
  (b) Network size $n = 800$.\\
  \includegraphics[width=\textwidth]{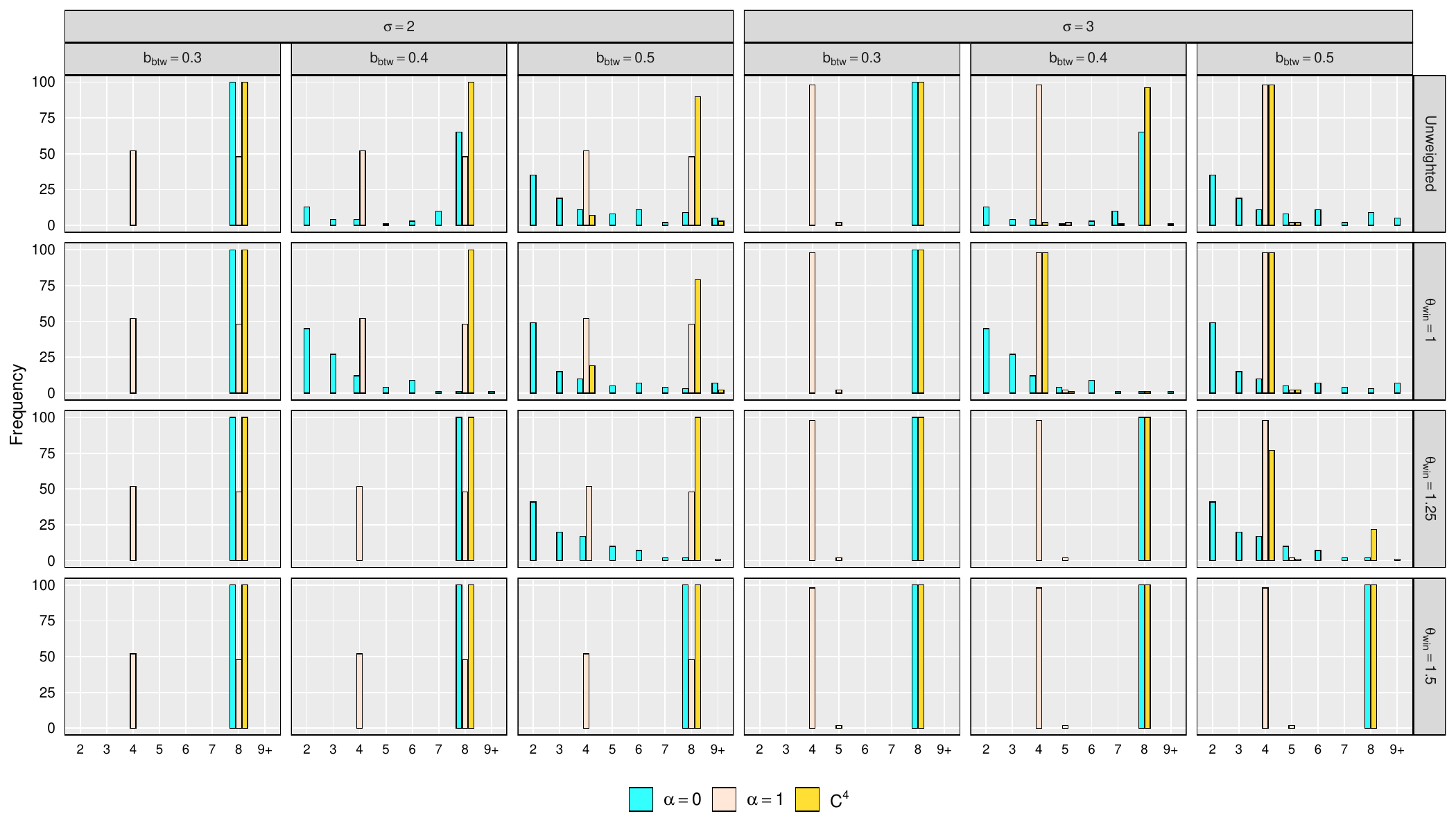}
  \caption{Distribution of the $K$ selection  under 
  $K_{\text{true}}=8$ among all simulation configurations with 
  $n=400$ (upper) and $n=800$ (lower).}
  \label{fig:K8_dist}
\end{figure}

Figure~\ref{fig:K4_dist} and~\ref{fig:K8_dist} present the empirical 
distributions of the estimated $K$ for $\alpha = 0$, $\alpha = 1$, 
and $\text{C}^4$, respectively, across all simulation configurations. 
It is worth noting that in extreme scenarios where structural signal 
is severely weak and covariate signal is not strong either (e.g. 
$b_{\text{btw}} = 0.5$, $K = 8$, $\sigma = 3$, and 
$n \in \{400, 800\}$, with $\theta_{\text{win}} = 1$ or the 
unweighted network, corresponding to the first and second rows in the 
last columns of both panels in Figure~\ref{fig:K8_dist}.), 
the distribution of $K$ selection for $\text{C}^4$ becomes identical 
to $\sigma = 1$. In these scenarios, $\text{C}^4$ automatically 
detects that structural information is totally 
unreliable as noise and selects $\alpha$ to be $1$ as the estimated 
tuning parameter. Although $\alpha = 0$ method occasionally estimates 
the correct $K = 8$ in these extreme settings, 
cross-comparison with Figure~\ref{fig:unknown_box_n} shows that these 
rare correct identifications still result in very poor clustering 
performance. Therefore, $\text{C}^4$ which relies on covariates only 
in these scenarios represents a rational adaptation that prioritizes 
clustering quality over nominally correct but uninformative $K$ 
identification. However, in real 
applications, if such extreme scenario happens, users should 
carefully reassess the data and manually set $\alpha$ range based on 
domain knowledge to obtain more reliable results.

\bibliographystyle{chicago}
\bibliography{refs}

\end{document}